\documentclass[mypaper,8pt,twoside]{CoAst}
\usepackage{epsf,graphicx,fancyhdr,sfmath}
\renewcommand{\chaptermark}[1]{\markboth{#1}{}}

\def\kms {km~s$^{-1}$}

\newcommand{\kopf}{\small\itshape Comm.\ in Asteroseismology, N$^{\textsf{\underline{o}}}$ 159, 2009\\
Proceedings of the JENAM 2008 Symposium N$^{\textsf{\underline{o}}}$~4:
Asteroseismology and Stellar Evolution}

\newcommand{\Authors}[1]{\begin{center}\normalsize\bf\sf #1 \end{center}}

\renewcommand{\author}[1]{\begin{center}\normalsize\bf\sf #1 \end{center}}
\newcommand{\Address}[1]{\begin{center}\small\sf #1 \end{center}}

\newcommand{\Objects}[1]{{\vspace{3mm}\small \noindent Individual Objects: }\small\sf \hangindent=27truemm \hangafter=1 #1 }

\renewenvironment{abstract}{\section*{Abstract}\normalsize\sf}{}
\newcommand{\References}[1]{\begin{flushleft}{\large References\\}\vspace*{2mm}\small #1 \end{flushleft}}

\newcommand{\chapterCoAst}[2]{\chapter[\sf\normalsize #1\\ \footnotesize \hspace*{5mm}by #2 \sf\normalsize][]{#1\\}\rhead[\fancyplain{}{\sf\footnotesize \center{#1}}]{\fancyplain{}{\sffamily\thepage}}\lhead[\fancyplain{\kopf}{\sffamily\thepage}]{\fancyplain{\kopf}{\sf\footnotesize \center{#2}}}}

\renewcommand{\chaptername}{}

\def\rfr{\smallskip\par\noindent
        \hangindent=7truemm
        \hangafter=1}

\def\ms{m\,s$^{-1}$}
\def\kms{km\,s$^{-1}$}
\def\boo{$\lambda$\,Boo}
\def\scu{$\delta$\,Scuti}
\def\cep{$\beta$\,Cephei}

\begin{document}
\sf

\chapterCoAst{Asteroseismology of chemically peculiar stars}
{O.\,Kochukhov} 
\Authors{O.\,Kochukhov} 
\Address{Department of Physics and Astronomy, Uppsala University, \\Box 515,SE-751 20 Uppsala, Sweden}

\textit{The section ``Rapidly oscillating magnetic Ap stars'' of this review is an updated version of the
paper published in the proceedings of the Wroc{\l}aw HELAS Workshop ``Interpretation of Asteroseismic Data'',
CoAst, 157, in press, arXiv:0810.1508}

\noindent
\begin{abstract}
Pulsational variability is observed in several types of main sequence stars
with anomalous chemical abundances. In this contribution I summarize the relationship between
pulsations and chemical peculiarities, giving special emphasis to
rapid oscillations in magnetic Ap stars. These magneto-acoustic pulsators provide
unique opportunities to study the
interaction of pulsations, chemical inhomogeneities, and strong magnetic
fields. Time-series monitoring of rapidly oscillating Ap stars using
high-resolution spectrometers at large telescopes and ultra-precise space
photometry has led to a number of important breakthroughs in our understanding
of these interesting objects. Interpretation of the roAp frequency spectra has
allowed constraining fundamental stellar parameters and probing poorly known
properties of the stellar interiors. At the
same time, investigation of the pulsational wave propagation in chemically
stratified atmospheres of roAp stars has been used as a novel asteroseismic
tool to study pulsations as a function of atmospheric height and to map in
detail the horizontal structure of the magnetically-distorted {\it p-}modes. 
\end{abstract}

\Objects{HR\,8799, HD\,116114, HD\,201601 ($\gamma$\,Equ), HD\,176232 (10\,Aql), HD\,134214, HD\,137949
(33\,Lib), HD\,99563, HD\,24712, HD\,75445, HD\,137909 ($\beta$\,CrB), HD\,101065, HR\,3831.}

\section*{Introduction}

On and near the main sequence, for spectral types from B to early F, one finds a remarkable
diversity of the stellar surface properties and variability. In cooler and hotter parts of the
H-R diagram a single, powerful process, such as convection in solar-type stars or mass loss in
hot massive stars, dominates the physics of stellar atmospheres. In contrast, several processes
of comparable magnitude compete in the A-star atmospheres and envelopes, creating interesting and
heterogeneous stellar population. The radiative diffusion (Michaud 1970) is the most important
process responsible for non-solar surface chemical composition. The diffusion theory suggests
that ions heavier than hydrogen are able to levitate or sink under competing influence of the
radiation pressure and gravity. Element segregation by the radiative diffusion is easily wiped
out by various hydrodynamical mixing effects and, thus, requires a star which is stable over
significant part of its outer envelope. Slowly rotating B-F stars with shallow convection zones
provide the required stability. The presence of strong, global magnetic field contributes further
to the suppression of turbulence and leads to different diffusion velocities depending on the
field inclination and strength. Chemically peculiar stars are separated into the two distinct
sequences according to their magnetic properties. Am, \boo, HgMn stars lack strong
magnetic fields and show mild chemical anomalies in chemically homogeneous outer stellar layers.
Ap and Bp stars have magnetic fields exceeding few hundred gauss, exhibit extreme chemical
anomalies and have substantial vertical and horizontal chemical gradients in the photosphere.

Stellar variability, including pulsations, adds important time-dependent aspect to the complex
picture of chemically peculiar B--F stars. Depending on the pulsation frequency and the physics
of the interaction
between mode excitation, composition gradients and magnetic field, different types of pulsations
are suppressed or excited. Observation and asteroseismic interpretation of this pulsational
variability 
is a powerful tool for determining fundamental stellar
parameters and constraining poorly known interior properties of chemically peculiar stars.

In this review I summarize our current understanding of the relationship between stellar
pulsations and chemical peculiarity for stars in the roAp, \scu, SPB and \cep\ instability
regions.

\section*{Metallic line A stars}

There is near-exclusion of \scu\ pulsations and Am-type chemical peculiarity (e.g., Breger 1970).
The diffusion theory explains this by invoking gravitational settling, which 
removes He from the He~{\sc ii} ionization zone, suppressing the driving of \scu\ pulsations.
When the Am star evolves off the main sequence, the He~{\sc ii} ionization region shifts deeper
into the star and reaches layers where some residual He is left. This allows excitation of 
low-amplitude \scu\ pulsations (Cox et al. 1979). These evolved \scu\ variables with residual
Am-like chemical peculiarities are known as $\rho$~Pup stars.

Classical \scu\ pulsations have been also claimed in several unevolved Am and Ap stars (e.g.,
Kurtz 1989; Martinez et al. 1999; Gonz\'alez et al. 2008). In some of these cases the detection
of pulsational variability is convincing. However, the Ap or Am nature of the stars in question,
inferred from photometry and old low-resolution classification spectra, is very uncertain.
Detailed model atmosphere and abundance analyses using modern, high-quality spectroscopic
material are required to confirm or refute the suspected unusual combination of high-amplitude
\scu\ pulsation and large chemical peculiarities.

\section*{Pulsating $\lambda$~Bootis stars}

\boo\ stars are Population I early-A to early-F type stars which exhibit significant
underabundance of most iron-peak and heavy elements but show solar abundances of CNO and
some other light elements (Paunzen et al. 2002a; Heiter 2002). These chemical properties
are believed to arise from contamination of the shallow stellar surface convection zones by
the accretion of metal-depleted gas from a circumstellar shell (Venn \& Lambert 1990) or
a diffuse interstellar cloud (Kamp \& Paunzen 2002).

The H-R diagram position of the \boo\ group members partially overlaps with the \scu\
instability strip. While accumulation of metals and He depletion prevents \scu\ type
pulsations in most Ap and Am stars, the opposite abundance signatures of \boo\ stars make
them more promising targets for pulsational observations. In particular, asteroseismic
investigations of these stars are interesting for constraining the stellar fundamental
parameters and determining the average metal content of the stellar interiors.

High-resolution time-series spectroscopy by Bohlender et al. (1999) revealed the presence
of high-degree non-radial pulsations in the majority of investigated \boo\ stars. The
overall pulsational characteristics of the group were summarized by Paunzen et al. (2002b).
They concluded that the fraction of pulsating \boo\ stars inside the \scu\ instability
strip (at least 70\%) is significantly larger than for normal stars. Moreover, in contrast
to classical \scu\ stars, which often pulsate in the fundamental mode, \boo\ stars tend to
pulsate in high-overtone modes.

Interestingly, at least one object with \boo\ chemical characteristics -- the planetary
host star HR\,8799 -- is known to exhibit the $\gamma$~Dor type pulsational variability
(Zerbi et al. 1999; Gray \& Kaye 1999). However, HR\,8799 appears to be an exception as
other members of the $\gamma$~Dor group show normal abundance pattern (Bruntt et al. 2008).

\section*{Pulsations and chemical peculiarity in hot stars}

The hot pulsating stars (Slowly Pulsating B and \cep), chemically peculiar Bp stars, and
non-pulsating normal B stars coexist in the same part of the H-R diagram (Briquet et al. 2007).
Nevertheless, up to now no conclusive evidence for the significant overlap of the pulsational,
magnetic, and chemical peculiarity phenomena has been identified for B-type stars. Analysis of
the low-resolution UV flux distributions showed that metallicities of SPB (Niemczura 2003) and
\cep\ (Niemczura \&  Daszynska-Daszkiewicz 2005) pulsators do not differ from those of normal B
stars. On the other hand, Morel et al. (2008) suggested the existence of a population of
nitrogen-rich and boron-depleted slowly rotating B stars based on NLTE abundance analysis of
high-resolution spectra. It is possible that the photospheric chemistry of these objects is
altered by a weak magnetic field in qualitatively the same way as much stronger fields of Bp
stars lead to prominent deviations from the solar chemical composition. However, apart from a
small number of SPB and \cep\ stars with $\sim$\,100~G fields (Neiner et al. 2003; Hubrig et al.
2006a), the universal presence of weak magnetic fields could not be convincingly established for
normal and/or pulsating B-type stars.

The non-magnetic HgMn chemically peculiar stars present another challenge for our understanding
of the excitation of pulsations in hot stars. Many HgMn stars are situated within the SPB
instability strip. Furthermore, an increased opacity due to accumulation of metals by radiative
diffusion in HgMn stars is expected to enhance the driving of the SPB pulsations (Turcotte \&
Richard 2002). Contrary to this theoretical prediction photometric observations show no evidence
of pulsational variability in HgMn stars (Adelman 1998). Spectroscopic line profile
variations detected for a handful of HgMn stars is limited to lines of 2--3 heavy elements and is,
consequently, attributed to chemical inhomogeneities rather than pulsation (Adelman et al. 2002;
Kochukhov et al. 2005; Hubrig et al. 2006b). Incompleteness of the theoretical diffusion models
in the outer part of the stellar envelope is the most likely explanation for the contradiction
between predicted and observed pulsation properties of HgMn stars.

\section*{Rapidly oscillating magnetic Ap stars}

This section is an updated version of the review published by Kochukhov (2008, 
in the proceedings of the Wroc{\l}aw HELAS Workshop ``Interpretation of Asteroseismic Data'',
CoAst, 157, in press).

Rapidly oscillating Ap (roAp) stars represent the most prominent subgroup of pulsating chemically
peculiar stars. These objects belong to the SrCrEu type of magnetic A stars, and pulsate in
high-overtone, low degree {\it p-}modes. roAp stars are found at or near the main sequence, at
the cool border of the region occupied by the magnetic Ap/Bp stars (Kochukhov \& Bagnulo 2006).
According to the series of recent spectroscopic  studies (e.g., Ryabchikova et al. 2002, 2004;
Kochukhov et al. 2002a), effective temperatures of roAp stars range from about 8100 down to 6400~K. Their
atmospheres are characterized by diverse chemical abundance patterns, but typically have normal
or below solar concentration of light and iron-peak elements and a very large overabundance of
rare-earth elements (REEs). Similar to other cool magnetic A  stars, roAp stars possess global
fields with a typical strength from few to ten kG (Mathys et al. 1997), although in some stars
the field intensity can exceed 20~kG (Kurtz et al. 2006b). These global magnetic topologies are
most likely the remnants of the fields which were swept at the star-formation phase or generated
by dynamo in the convective envelopes of pre-main sequence stars, then quickly decayed to a
stable configuration (Braithwaite \& Nordlund 2006) and now remain nearly constant on stellar
evolutionary timescales. The slow rotation and stabilizing effect of the strong magnetic field
facilitates operation of the atomic diffusion processes (Michaud et al. 1981; LeBlanc \& Monin
2004), which are responsible for the  grossly non-solar surface chemistry and large element
concentration gradients in Ap-star atmospheres (Ryabchikova et al. 2002, 2008; Kochukhov et al.
2006). Variation of the field strength and inclination across the stellar surface alters the
local diffusion velocities (Alecian \& Stift 2006), leading to the formation of spotted chemical
distributions and consequential synchronous rotational modulation of the broad-band photometric
indices, spectral line profiles, the longitudinal magnetic field and the mean field modulus (e.g.,
Ryabchikova et al. 1997).

Pulsations in cool Ap stars were discovered 30 years ago (Kurtz 1978) and were immediately
recognized to be another manifestation of the prominent influence of  unusually strong magnetic
fields on the stellar interiors and atmospheres.  Currently, 38 cool Ap stars are known to
pulsate. Several new roAp stars were recently discovered using high-resolution spectroscopic observations
(Hatzes \& Mkrtichian 2004; Elkin et al. 2005; Kurtz et al. 2006b; Kochukhov et al. 2008a, 2009;
Gonz\'ales et al. 2008). Oscillations have amplitudes below 10 mmag in the Johnson's B filter and
0.05--5~\kms\ in spectroscopy, while the periods lie in the range from 4 to 22~min. The latter
upper period threshold of roAp pulsation corresponds  to the second mode recently detected
by the high-precision RV observations of the evolved Ap star HD\,116114 (Kochukhov, Bagnulo \& Lo
Curto, in preparation). 

The amplitude and phase of
pulsational variability are modulated with the stellar rotation. A simple geometrical
interpretation of this phenomenon was suggested by the oblique pulsator model of Kurtz (1982),
which supposes an alignment of the low angular degree modes with the quasi-dipolar magnetic
field of the star and resulting variation of the aspect at which pulsations are seen by the
distant observer. Detailed theoretical studies (Bigot \& Dziembowski 2002; Saio 2005) showed
that the horizontal geometry of {\it p-}mode pulsations in magnetic stars is far
more complicated: individual modes are distorted by the magnetic field and rotation in such a
way  that pulsational perturbation cannot be approximated by a single spherical harmonic
function.

\subsection*{Photometric studies of roAp pulsations}

Majority of roAp stars were discovered by D. Kurtz and collaborators using photometric
observations at SAAO (see review by Kurtz \& Martinez 2000). The search for roAp stars in the
Northern hemisphere is being conducted at the Nainital (Joshi et al. 2006) and Maidanak
(Dorokhova \& Dorokhov 2005) observatories. Several roAp stars were  observed in coordinated
multi-site  photometric campaigns (Kurtz et al. 2005a; Handler et al. 2006), which allowed to
deduce frequencies with the precision sufficient for asteroseismic analysis. However, low
amplitudes of broad-band photometric variation of roAp stars, low duty cycle and aliasing
problems inevitably limit precision of the ground-based photometry. Instead of pursuing
observations from the ground, recent significant progress has been achieved by uninterrupted,
ultra-high precision observations of known roAp stars using small photometric telescopes in
space. Here the Canadian MOST space telescope is undisputed leader. The MOST team has completed
3--4 week runs on HD~24712, $\gamma$~Equ (HD~201601), 10~Aql (HD~176232),  HD~134214, and
HD~99563. Observations of 33~Lib (HD~137949) are planned for April-May 2009.

In addition to providing unique material for detailed asteroseismic studies of HD~24712,
$\gamma$\,Equ, and 10\,Aql, the MOST photometry has revealed the presence of a very close
frequency pair in $\gamma$~Equ, giving modulation of pulsation amplitude with $\approx$18~d
period (Huber et al. 2008). It is possible that this frequency beating is responsible for the
puzzling discrepancy of the radial velocity amplitudes found for $\gamma$~Equ in different short
spectroscopic observing runs  (Sachkov et al. 2009). This amplitude variation could not be
ascribed to the rotational modulation because rotation period of this star exceeds 70 years
(Bychkov et al. 2006).

\subsection*{Spectroscopy of roAp pulsations}

High-quality time-resolved spectra of roAp stars have proven to be the source of new, incredibly
rich information, which not only opened new possibilities for the research on magneto-acoustic
pulsations but yielded results of wide astrophysical significance. Numerous spectroscopic
studies of individual roAp stars (e.g., Kochukhov \& Ryabchikova 2001a; Mkrtichian et al. 2003;
Ryabchikova et al. 2007a), as well as comprehensive analysis of pulsational variability in 10
roAp stars published by Ryabchikova et al. (2007b), demonstrated pulsations in spectral lines
very different from those observed in any other type  of non-radially pulsating stars. The most
prominent characteristic of the RV oscillation in roAp stars is the extreme diversity of
pulsation signatures of different elements. Only a few stars show evidence of $<$50~\ms\
variation in the lines of iron-peak elements, whereas REE lines, especially those of Nd~{\sc
ii}, Nd~{\sc iii}, Pr~{\sc iii},  Dy~{\sc iii}, and Tb~{\sc iii} exhibit amplitudes  ranging
from a few hundred \ms\ to several \kms. The narrow core of H$\alpha$ behaves similarly to REE
lines (Kochukhov 2003; Ryabchikova et al. 2007b), suggesting line formation at comparable
atmospheric heights.

Pulsation phase also changes significantly from one line to another (Kochukhov \& Ryabchikova
2001a; Mkrtichian et al. 2003), with the most notorious example of 33~Lib where different
lines of {\it the same ion} pulsate with a 180$^{\rm o}$ shift in phase, revealing a radial
node, and show very different ratios of the amplitude at the main frequency and its
first harmonic (Ryabchikova et al. 2007b). Several studies concluded that, in general, roAp
stars show a combination of running (changing phase) and
standing (constant phase) pulsation waves at different atmospheric heights.

Another unusual aspect of the spectroscopic pulsations in roAp stars is a large change of
the oscillation amplitude and phase from the line core to the wings. The bisector variation expected
for the regular spherical harmonic oscillation is unremarkable and should exhibit neither
changing phase nor significantly varying amplitude. Contrary to this expectation of the
common single-layer pulsation model, the roAp bisector 
amplitude often shows an increase from 200--400~\ms\ in the cores of strong REE lines to
2--3~\kms\ in the line wings, accompanied by significant changes of the bisector phase (Sachkov
et al. 2004; Kurtz et al. 2005b; Ryabchikova et al. 2007b).

The ability to resolve and measure with high precision pulsational variation in individual lines
allows to focus analysis on the spectral features most sensitive to pulsations. By co-adding radial
velocity curves of many REE lines recorded in a spectrum with a wide wavelength coverage 
one is able to reach the RV accuracy of $\sim$\,1~\ms\ (Mathys et al. 2007). This
made possible discovery of the very low-amplitude oscillations in HD\,75445 (Kochukhov et al.
2009) and HD\,137909 (Hatzes \& Mkrtichian 2004). The second object, well-known cool Ap star
$\beta$~CrB, was previously considered to be a typical non-pulsating Ap (noAp) star due to 
null results of numerous photometric searches of pulsations (Martinez \& Kurtz 1994) and the
absence of prominent REE ionization anomaly found for nearly all other roAp stars (Ryabchikova et
al. 2001, 2004). The fact that $\beta$~CrB is revealed as the second brightest roAp star
corroborates the idea that {\it p-}mode oscillations could be present in all cool Ap stars
but low pulsation amplitudes prevented detection of pulsations in the so-called noAp stars
(Kochukhov et al. 2002b; Ryabchikova et al. 2004).

Despite the improved sensitivity in searches of the low-amplitude oscillations in roAp candidates
and numerous outstanding discoveries for known roAp stars, the major limitation of the
high-resolution spectroscopic monitoring is a relatively small amount of observing time
available at large telescopes for these projects. As a result, only snapshot time-series spanning
2--4 hours were recorded for most roAp stars, thus providing an incomplete and, possibly, biased
picture for the multiperiodic pulsators, for which close frequencies cannot be resolved in such
short runs. Observations on different nights, required to infer detailed RV frequency spectrum,
were secured only for a few roAp stars (Kochukhov 2006; Mkrtichian et al. 2008). For example, in
recent multi-site spectroscopic campaign carried out for 10~Aql using two telescopes on seven
observing nights (Sachkov et al. 2008), we found that beating of the three dominant
frequencies leads to strong changes of the apparent RV amplitude during several hours. This
phenomenon could explain puzzling modulation of the RV pulsations on timescales of 1--2 hours
detected in  some roAp stars (Kochukhov \& Ryabchikova 2001b; Kurtz et al. 2006a).

\subsection*{Asteroseismology of roAp stars}

The question of the roAp excitation mechanism has been debated for many years but now is narrowed
down to the $\kappa$ mechanism acting in the hydrogen ionization zone, with the additional
influence from the magnetic quenching of convection and composition gradients built up by the
atomic diffusion (Balmforth et al. 2001; Cunha 2002; Vauclair \& Th\'eado 2004). However,
theories cannot reproduce the observed temperature and luminosity distribution of roAp stars and
have not been able to identify parameters distinguishing pulsating Ap stars from their apparently
constant, but otherwise very similar, counterparts (Th\'eado et al. 2009). At the same time, some
success has been achieved in calculating magnetic perturbation of oscillation frequencies (Cunha
\& Gough 2000; Saio \& Gautschy 2004) and inferring fundamental parameters and  interior
properties for multiperiodic roAp stars (Matthews et al. 1999; Cunha et al. 2003).

Recent asteroseismic interpretation of the frequencies deduced
from the  MOST data for $\gamma$~Equ (Gruberbauer et al. 2008) and 10~Aql (Huber et al. 2008)
yields stellar parameters in good agreement with detailed model
atmosphere studies. At the same time, the magnetic field required by seismic models to
fit the observed frequencies is 2--3 times stronger than the field modulus inferred from the Zeeman
split spectral lines. This discrepancy could be an indication that magnetic field in the
{\it p-}mode driving zone is significantly stronger than the surface field or it may 
reflect an incompleteness of the theoretical models.

Mkrtichian et al. (2008) presented the first detailed asteroseismic analysis of a roAp star based
entirely on spectroscopic observations. Using high-precision RV measurements spanning four
consecutive nights, the authors detected 26 frequencies for famous roAp star HD\,101065
(Przybylski's star). Mode identification showed the presence of 15 individual modes with
$\ell$\,=\,0--2. This rich frequency spectrum of HD\,101065 can be well reproduced by theoretical
models if an excessively strong ($\approx$\,9~kG) dipolar magnetic field is assumed, in
contradiction to  $\langle B \rangle$\,=\,2.3~kG inferred directly from the stellar spectrum
(Cowley et al. 2000).

\subsection*{Tomography of atmospheric pulsations in roAp stars}

The key observational signature of roAp pulsations in spectroscopy -- large line-to-line
variation of pulsation amplitude and phase -- is understood in terms of an interplay between
pulsations and chemical stratification. The studies by Ryabchikova et al. (2002, 2008) and
Kochukhov et al. (2006) demonstrated that light and iron-peak elements tend to be
overabundant in deep atmospheric layers (typically $\log\tau_{5000}\ge -0.5$) of cool Ap
stars, which agrees with the predictions of self-consistent diffusion models (LeBlanc \&
Monin 2004). On the other hand, REEs accumulate in a cloud at very low optical depth. The NLTE
stratification studies, performed for Nd and Pr ions, place the lower boundary of this cloud
at $\log\tau_{5000}\approx -3$ (Mashonkina et al. 2005, 2009). Then, the rise of pulsation amplitude
towards the upper atmospheric layers due to exponential density decrease does not affect Ca,
Fe, and Cr lines but shows up prominently in the core of H$\alpha$ and in REE lines. This
picture of the pulsation waves propagating outwards through the stellar atmosphere with
highly inhomogeneous chemistry has gained general support from observations and theoretical
studies alike. Hence the properties of roAp atmospheres allow an entirely new type of
asteroseismic analysis -- vertical resolution of {\it p-}mode cross-sections simultaneously
with the constraints on distribution of chemical abundances. 

The two complimentary approaches to the roAp pulsation tomography problem have been discussed by
Ryabchikova et al. (2007a, 2007b). On the one hand, tedious and detailed line formation
calculations, including stratification analysis, NLTE line formation, sophisticated model
atmospheres and polarized radiative transfer, can supply mean formation heights for
individual pulsating lines. Then, the pulsation mode structure can be mapped directly by
plotting pulsation amplitude and phase of selected lines against optical or geometrical depth.
On the other hand, the phase-amplitude diagram method proposed by Ryabchikova et al. (2007b)
is suitable for a coarse analysis of the vertical pulsation structure without invoking model
atmosphere calculations but assuming the presence of the outwardly propagating wave
characterized by a continuous change of amplitude and phase. In this case, a scatter plot of the
RV measurements in the phase-amplitude plane can be interpreted in terms of the standing and
running waves, propagating in different parts of the atmosphere.

To learn about the physics of roAp atmospheric oscillations one should compare
empirical pulsation maps with theoretical models of the {\it p-}mode
propagation in magnetically-dominant ($\beta<<1$) part of the stellar envelope. Sousa \& Cunha (2008)
considered an analytical model of the radial modes in an isothermal atmosphere with
exponential density decrease. They argue that waves are decoupled into the standing magnetic
and running acoustic components, oriented perpendicular and along the magnetic field lines,
respectively. The total projected pulsation velocity, produced by a superposition of these
two components, can have widely different vertical profile depending on the magnetic field
strength, inclination and the aspect angle. For certain magnetic field parameters and viewing
geometries the two  components cancel out, creating a node-like structure. This model
can possibly account for observations of radial nodes in 33~Lib (Mkrtichian et al.
2003)  and 10~Aql (Sachkov et al. 2008). 

The question of interpreting the line profile variation (LPV) of roAp stars has recieved great
attention after it was demonstrated that the REE lines in $\gamma$~Equ exhibit unusual
blue-to-red asymmetric variation (Kochukhov \& Ryabchikova 2001a), which is entirely  unexpected
for a slowly rotating non-radial pulsator. Kochukhov et al. (2007) showed the presence of similar
LPV in the REE lines of several other roAp stars and presented examples of the transformation
from the usual symmetric blue-red-blue LPV in Nd~{\sc ii} lines to the asymmetric blue-to-red
waves in the Pr~{\sc iii} and Dy~{\sc iii} lines formed higher in the atmosphere. These lines
also show anomalously broad profiles (e.g., Ryabchikova et al. 2007b), suggesting existence of an
isotropic velocity field, with dispersion of the order of 10~\kms, in the uppermost atmospheric layers. Kochukhov
et al. (2007) proposed a phenomenological model of the interaction between this turbulent layer
and pulsations that has successfully reproduced asymmetric LPV of doubly ionized REE lines. An
alternative model by Shibahashi et al. (2008) obtains similar LPV by postulating formation of REE
lines at extremely low optical depths, in disagreement with the detailed NLTE calculations by
Mashonkina et al. (2005, 2009), and requires the presence of shock waves in stellar atmospheres,
which is impossible to reconcile with the fact that observed RV amplitudes are well below the
sound speed.

Oblique pulsations and distortion of non-radial modes by rotation and magnetic field precludes
direct application of the standard mode identification techniques to roAp stars. A meaningful
study of their horizontal pulsation geometry became possible by using the method of pulsation
Doppler imaging (Kochukhov 2004a). This technique derives maps of pulsational fluctuations
without making {\it a priori} assumption of the spherical harmonic pulsation geometry.
Application of this method to HR\,3831 (Kochukhov 2004b) provided the first independent
verification of the oblique pulsator model by showing alignment of the axisymmetric pulsations
with the symmetry axis of the stellar magnetic field. At the same time, Saio (2005) showed that
the observed deviation of the oscillation geometry of HR\,3831 from a oblique dipole mode agrees
well with his model of magnetically distorted pulsation.

\section*{Outlook}

A progress in understanding the relation between the phenomena of chemical peculiarity and stellar pulsations
calls for a detailed model atmosphere and chemical abundance analysis of the suspected high-amplitude \scu\
Ap and classical Am stars. Interpretation of the modern high-resolution spectroscopic observational material
using realistic model atmospheres is also needed to clarify the question of the connection between CP and hot
pulsating stars. Systematic high-resolution spectropolarimetric observations are urgently needed to verify
the claims of weak magnetic fields in many SPB and a few \cep\ stars. On the other hand, comprehensive
theoretical modelling is needed to explore asteroseismic potential of the pulsating \boo\ stars and, in
particular, to test possibilities of constraining their interior chemical profiles.

For roAp stars, several important open questions and promising research directions can be identified. On the
theoretical side, the failure of the current pulsation models to account for the observed blue and red
borders of the roAp instability strip should be addressed by including a more realistic physical description
of the interplay between pulsations, magnetic fields, stratified chemistry, and stellar rotation. On the
observational side, systematic spectroscopic searches for low-amplitude magnetoacoustic oscillations in cool
Ap stars are evidently needed to overcome the limitations and biases of previous photometric surveys.

The remarkable spectroscopic pulsational behaviour, demonstrated in numerous recent studies of roAp stars,
extends the roAp research to the uncharted territory far beyond the field of classical asteroseismology. In
addition to interpretation of pulsation frequencies, roAp stars now offer a unique opportunity for
\textit{pulsation tomography}, i.e. a study of different pulsation modes in 3-D, made possible by the
rotational modulation of the oblique pulsations and  a prominent effect of chemical stratification.
Spectacular observational results, such as resolution of the vertical pulsation mode cross-sections and
Doppler imaging of atmospheric pulsations in roAp stars, are, however, yet to be matched by corresponding
theoretical developments. At the moment we lack realistic models treating propagation of pulsation waves in
the outer layers of magnetic Ap stars. Our knowledge about chemical stratification, in particular that of
rare-earth elements, and its impact on the atmospheric structure is equally incomplete. Addressing these
theoretical questions is required for the development of a solid physical basis for astrophysical
interpretation of the recent roAp pulsation tomography results.

\References{
\rfr Adelman S.J. 1998, A\&ASS, 132, 93
\rfr Adelman S.J., Gulliver A.F., Kochukhov O.P., \& Ryabchikova T.A. 2002, ApJ, 575, 449
\rfr Alecian G., \& Stift M.J. 2006, A\&A, 454, 571
\rfr Balmforth N.J., Cunha M.S, Dolez N., et al. 2001, MNRAS, 323, 362
\rfr Bigot L., \& Dziembowski W.A. 2002, A\&A, 391, 235
\rfr Bohlender D.A., Gonz\'alez J.F., \& Matthews J.M. 1999, A\&A, 350, 553
\rfr Braithwaite J., \& Nordlund \AA. 2006, A\&A, 450, 1077
\rfr Breger M. 1970, ApJ, 162, 597
\rfr Briquet M., Hubrig S., De Cat P., et al. 2007, A\&A, 466, 269
\rfr Bruntt H., De Cat P., \& Aerts C. 2008, A\&A, 478, 487
\rfr Bychkov V.D., Bychkova L.V., \& Madej J. 2006, MNRAS, 365, 585
\rfr Cowley C.R., Ryabchikova T., Kupka F., et al. 2000, MNRAS, 317, 299
\rfr Cox A.N., King D.S., \& Hodson S.W. 1979, ApJ, 231, 798
\rfr Cunha M.S., \& Gough D. 2000, MNRAS, 319, 1020
\rfr Cunha M.S. 2002, MNRAS, 333, 47
\rfr Cunha M.S., Fernandes J.M.M.B., \& Monteiro, M.J.P.F.G. 2003, MNRAS, 343, 831  
\rfr Dorokhova T., \& Dorokhov N. 2005, JApA, 26 223
\rfr Elkin V.G., Rilej J., Cunha M., et al. 2005, MNRAS, 358, 665
\rfr Gonz\'alez J.F, Hubrig S., Kurtz D.W., Elkin V., \& Savanov I. 2008, 384, 1140
\rfr Gruberbauer M., Saio H., Huber D., et al. 2008, A\&A, 480, 223
\rfr Gray R.O, \& Kaye A.B. 1999, AJ, 118, 2993
\rfr Handler G., Weiss W.W., Shobbrook R.R., et al. 2006, MNRAS, 366, 257
\rfr Hatzes A.P., \& Mkrtichian D.E. 2004, MNRAS, 351, 663
\rfr Heiter U. 2002, A\&A, 381, 959
\rfr Huber D., Saio H., Gruberbauer M., et al. 2008, A\&A, 483, 239
\rfr Hubrig S., Briquet M., Sch\"oller M., et al. 2006a, MNRAS, 369, L61
\rfr Hubrig S., Gonz\'alez J.F., Savanov I., et al. 2006b, MNRAS, 371, 1953
\rfr Joshi S., Mary D.L., Martinez P., et al. 2006, A\&A, 455, 303
\rfr Kamp I., \& Paunzen E. 2002, MNRAS, 335, L45
\rfr Kochukhov O., \& Ryabchikova T. 2001a, A\&A, 374, 615
\rfr Kochukhov O., \& Ryabchikova T. 2001b, A\&A, 377, L22
\rfr Kochukhov O., Bagnulo S., \& Barklem P.S. 2002, ApJ, 578, L75
\rfr Kochukhov O., Landstreet J.D., Ryabchikova T., Weiss W.W., \& Kupka F. 2002b, MNRAS, 337, L1
\rfr Kochukhov O. 2003, in {\it Magnetic Fields in O, B and A stars}, eds.
Balona L.A., Henrichs H.F., \& Medupe R., ASP Conf. Ser., 305, 104
\rfr Kochukhov O. 2004a, A\&A, 423, 613
\rfr Kochukhov O. 2004b, ApJ, 615, L149
\rfr Kochukhov O., Piskunov N., Sachkov M., \& Kudryavtsev D. 2005, A\&A, 439, 1093 
\rfr Kochukhov O. 2006, A\&A, 446, 1051
\rfr Kochukhov O. 2009, CoAst, 157, in press (arXiv:0810.1508)
\rfr Kochukhov O., \& Bagnulo S. 2006, A\&A, 450, 763
\rfr Kochukhov O., Bagnulo S., \& Barklem P.S. 2002, ApJ, 578, L75
\rfr Kochukhov O., Tsymbal V., Ryabchikova T., et al. 2006, A\&A, 460, 831
\rfr Kochukhov O., Ryabchikova T., Weiss W.W., et al. 2007, MNRAS, 376, 651
\rfr Kochukhov O., Ryabchikova T., Bagnulo S., \& Lo Curto G. 2008a, A\&A, 479, L29
\rfr Kochukhov O., Bagnulo S., Lo Curto G., \& Ryabchikova T., 2009, A\&A, in press (arXiv:0812.1565)
\rfr Kurtz D.W. 1978, IBVS, 1436
\rfr Kurtz D.W. 1982, MNRAS, 200, 807
\rfr Kurtz D.W. 1989, MNRAS, 238, 1077
\rfr Kurtz D.W., \& Martinez, P. 2000, Baltic Astronomy, 9, 253
\rfr Kurtz D.W., Elkin V.G., \& Mathys G. 2005a, MNRAS, 358, L10
\rfr Kurtz D.W., Cameron C., Cunha M.S., et al. 2005b, MNRAS, 358, 651
\rfr Kurtz D.W., Elkin V.G., \& Mathys G. 2006a, MNRAS, 370, 1274
\rfr Kurtz D.W., Elkin V.G., Cunha M.S., et al. 2006b, MNRAS, 372, 286
\rfr LeBlanc F., \& Monin D. 2004, in {\it IAU Symposium 224},
eds. {Zverko} J., {Ziznovsky} J., {Adelman} S.~J., \& {Weiss} W.~W., 193
\rfr Mashonkina L., Ryabchikova T., \& Ryabtsev V. 2005, A\&A, 441, 309
\rfr Mashonkina L., Ryabchikova T., Ryabtsev A., \& Kildiyarova R. 2009, A\&A, in press (arXiv:0811.3614)
\rfr Mathys G., Hubrig S., Landstreet J.D., et al. 1997, A\&AS, 123, 353
\rfr Mathys G., Kurtz D.W., \& Elkin V.G. 2007, MNRAS, 380, 181
\rfr Matthews J.M., Kurtz D.W., \& Martinez P. 1999, ApJ, 511, 422
\rfr Martinez P., \& Kurtz D.W. 1994, MNRAS, 271, 129
\rfr Martinez P., Kurtz D.W., \& Ashoka B.N., et al. 1999, MNRAS, 309, 871 
\rfr Michaud G. 1970, ApJ, 160, 641
\rfr Michaud G., Charland Y., \& Megessier C. 1981, A\&A, 103, 244
\rfr Mkrtichian D.E., Hatzes A.P., \& Kanaan A. 2003, MNRAS, 345, 781
\rfr Mkrtichian D.E., Hatzes A.P., Saio H., \& Shobbrook R.R. 2008, A\&A, 490, 1109
\rfr Morel T., Hubrig S., \& Briquet M. 2008, A\&A, 481, 453
\rfr Neiner C., Geers V.C., Henrichs H.F., et al. 2003, A\&A, 406, 1019
\rfr Niemczura E. 2003, A\&A, 404, 689
\rfr Niemczura E., \& Daszynska-Daszkiewicz J. 2005, A\&A, 433, 659
\rfr Paunzen E., Iliev I.Kh., Kamp I., \& Barzova I.S. 2002a, MNRAS, 336, 1030
\rfr Paunzen E., Handler G., Weiss W.W., et al. 2002b, A\&A, 392, 515
\rfr Ryabchikova T.A., Landstreet J.D., Gelbmann M.J., et al. 1997, A\&A, 327, 1137
\rfr Ryabchikova T.A., Savanov I.S., Malanushenko V.P., \& Kudryavtsev D.O. 2001, Astron.
Reports, 45, 382
\rfr Ryabchikova T., Piskunov N., Kochukhov O., et al. 2002, A\&A, 384, 545
\rfr Ryabchikova T., Nesvacil N., Weiss W.W., et al. 2004, A\&A, 423, 705
\rfr Ryabchikova T., Sachkov M., Weiss W.W., et al. 2007a, A\&A, 462, 1103
\rfr Ryabchikova T., Sachkov M., Kochukhov O., \& Lyashko D. 2007b, A\&A, 473, 907
\rfr Ryabchikova T., Kochukhov O., \& Bagnulo S. 2008, A\&A, 480, 811
\rfr Sachkov M., Ryabchikova T., Kochukhov O., et al. 2004, in {\it IAU Colloquium 193},
eds. Kurtz D.W., \& Pollard K.R., ASP Conf. Ser., 310, 208
\rfr Sachkov M., Kochukhov O., Ryabchikova T., et al. 2008, MNRAS, 389, 903
\rfr Sachkov M., Kochukhov O., Ryabchikova T., \& Gruberbauer M. 2009, in {\it Interpretation
of Asteroseismic Data}, CoAst, in press
\rfr Saio H., \& Gautschy A. 2004, MNRAS, 350, 485
\rfr Saio H. 2005, MNRAS, 360, 1022
\rfr Shibahashi H., Gough D., Kurtz D.W., \& Kambe E. 2008, PASJ, 60, 63
\rfr Sousa J., \& Cunha, M.S. 2008, CoSka, 38, 453
\rfr Th\'eado S., Dupret M.-A., Noels A., \& Ferguson J.W. 2009, A\&A, 493, 159
\rfr Turcotte S., \& Richard O. 2002, Ap\&SS, 284, 225
\rfr Vauclair S., \& Th\'eado S. 2004, A\&A, 425, 179
\rfr Venn K.A., \& Lambert D.L. 1990, ApJ, 363, 234
\rfr Zerbi F.M., Rodriguez E., Garrido R., et al. 1999, MNRAS, 303, 275
}

\end{document}